\documentclass[preprint,showpacs,preprintnumbers,amsmath,amssymb]{revtex4}
\usepackage{epsfig}
\usepackage{graphicx}
\usepackage{dcolumn}
\usepackage{bm}
\usepackage{threeparttable}

\def\beq{\begin{equation}}
\def\eeq{\end{equation}}
\def\eeqn{\end{equation}}
\newcommand\iden{\leavevmode\hbox{\small1\normalsize\kern-.33em1}}

\newcommand{\dd}{\mathrm{d}}

\newcommand{\bea} {\begin{eqnarray}}
\newcommand{\eea} {\end{eqnarray}}

\newcommand{\nn}{\nonumber}

\newcommand{\eqn}[1]{(\ref{#1})}
\newcommand{\nm}{\nonumber}

\let\jnfont=\rm
\def\NPB#1,{{\jnfont Nucl.\ Phys.\ B }{\bf #1},}
\def\PLB#1,{{\jnfont Phys.\ Lett.\ B }{\bf #1},}
\def\EPJC#1,{{\jnfont Eur.\ Phys.\ Jour.\ C }{\bf #1},}
\def\PRD#1,{{\jnfont Phys.\ Rev.\ D }{\bf #1},}
\def\PRL#1,{{\jnfont Phys.\ Rev.\ Lett.\ }{\bf #1},}
\def\MPLA#1,{{\jnfont Mod.\ Phys.\ Lett.\ A }{\bf #1},}
\def\JPG#1,{{\jnfont J.\ Phys.\ G }{\bf #1},}
\def\CTP#1,{{\jnfont Commun.\ Theor.\ Phys.\ }{\bf #1},}
\def\JHEP#1,{{\jnfont JHEP \ }{\bf #1},}
\def\NPPS#1,{{\jnfont Nucl.\ Phys.\ Proc.\ Suppl.\ }{\bf #1},}
\def\CPC#1,{{\jnfont Comput.\ Phys.\ Commun.\ }{\bf #1},}
\def\CPL#1,{{\jnfont Chin.\ Phys.\ Lett. }{\bf #1},}
\def\APPB#1,{{\jnfont Acta\ Phys.\ Polon.\ B }{\bf #1},}
\def\PR#1,{{\jnfont Phys.\ Rept.\  }{\bf #1},}
\def\CHC#1,{{\jnfont Chin.\ Phys.\ C }{\bf #1},}

\def\lsim{\raise0.3ex\hbox{$<$\kern-0.75em\raise-1.1ex\hbox{$\sim$}}}
\def\gsim{\raise0.3ex\hbox{$>$\kern-0.75em\raise-1.1ex\hbox{$\sim$}}}
\def\q_slash{\not{\hbox{\kern-2.1pt $q$}}}
\def\p_slash{\not{\hbox{\kern-2.1pt $p$}}}
\begin{document}

\title{\ \\[10mm]Muon g-2 Anomaly confronted with the higgs global data in the Left-Right Twin Higgs Models}

\author{Guo-Li Liu }\email{guoliliu@zzu.edu.cn}
\affiliation{ School of Physics, Zhengzhou University, ZhengZhou 450000, P. R. China }
\author{Qing-Guo Zeng }\email{zengqingguo66@126.com}
\affiliation{ School of Physics, Shangqiu Normal University, Shangqiu 470000, P. R. China }


\begin{abstract}
We will examine the Left-Right Twin Higgs(LRTH) Models as a solution of muon $g-2$ anomaly with the background of the Higgs global fit data.
In the calculation, the joint constrains from the theory, the precision electroweak data, the 125 GeV Higgs data,
the leptonic flavor changing decay $\mu \to e\gamma $ decays, and the constraints $m_{\nu_R}>m_T>m_{W_H}$ are all considered.
And with the small mass of the $\phi^0$, the direct searches from the $h\to \phi^0\phi^0$ channels can impose stringent
upper limits on Br$(h\to \phi^0\phi^0)$ and can reduce the allowed region of $m_{\phi^0}$ and $f$.
It is concluded that the muon g-2 anomaly can be explained in the region of 200 GeV $\leq M\leq$ 500 GeV,
700 GeV $\leq f\leq$ 1100 GeV, 13 GeV $\leq m_{\phi^0}\leq$ 55 GeV,  100 GeV $\leq m_{\phi^\pm}\leq$ 900 GeV, and $m_{\nu_R}\geq$ 15 TeV
after imposing all the constraints mentioned above.

\end{abstract}

\maketitle

\section{Introduction}
The muon anomalous magnetic moment ($g-2$) is a very precisely measured observable, and
expected to shed light on ¡±new physics¡±.
The muon $g-2$ anomaly has been
a long-standing puzzle since the announcement by the E821 experiment
in 2001 \cite{mug21,mug22}.
The precision measurement of $a_\mu=(g-2)/2$ has been performed by the E821 experiment at
Brookhaven National Laboratory \cite{BNL-g-2}, with the current
world-averaged result given by \cite{WAR-g-2}
\beq
a_\mu^{exp}= 116592091(\pm54)(\pm33)\times 10^{-11},
\eeq
Meanwhile, the Standard Model (SM) prediction from the Particle Data Group gives\cite{WAR-g-2},
\beq
a_\mu^{SM}= 116591803(\pm1)(\pm42)(\pm26)\times 10^{-11},
\eeq
The difference between experiment and theory is
\beq
\Delta a_\mu= a_\mu^{exp}-a_\mu^{SM}=(288\pm80)\times 10^{-11},
\eeq
which shows a $3.6\sigma$ discrepancy, hinting at tantalizing new
physics beyond the SM.
It is the difference between the experimental data and the SM prediction
determines the room for new physics.

There exist various new physics scenarios to explain the muon
$g-2$ excess, for recent reviews, see e.g.
Refs. \cite{g-2-review1,g-2-review2,g-2-review3,g-2-review4,g-2-review5}.
Among these extensions, the LRTH model may also provide a explanation for
the muon $g-2$ anomaly. In these models, there are six massive gauge bosons left after the symmetry breaking:
the SM $Z$ and $W^\pm$, and extra heavier bosons, $Z_{H}$ and $W_H^\pm$.
And these models also include eight scalars: one neutral pseudoscalar, $\phi^0$, a pair of charged
scalars $\phi^\pm$, the SM physical Higgs $h$, and an ${\rm SU}(2)_L$ twin Higgs
doublet $\hat{h}=(\hat{h}_1^+, \hat{h}_2^0)$.
The lepton couplings to the pseudoscalar can be sizably enhanced by the large right-handed neutrino mass $m_{\nu_R}$.
The pseudoscalar can give positive contributions to muon $g-2$ via the two-loop Barr-Zee diagrams.


In this work we will examine the parameter space of LRTH by considering the
joint constraints from the theory, the precision electroweak data,
 the 125 GeV Higgs signal data, the muon $g-2$ anomaly, the lepton rare decay of $\mu \to e \gamma$,
as well as the direct search limits from the LHC.

Our work is organized as follows. In Sec. II we recapitulate the
LRTH models. In Sec. III we discuss the muon $g-2$ anomaly and other relevant constraints. In Sec.
IV, we constrain the model using the direct search limits from the LHC, especially the Higgs global fit.
Finally, the conclusion is given in Sec. VI.

\section{The Relevant Couplings in the LRTH Models }
We need a global symmetry to implement the twin Higgs mechanism, and the global symmetry is partially
gauged and spontaneously broken. At the same time, to control
the quadratic divergences, we also need the twin symmetry which is identified with the left-right symmetry interchanging $L$
and $R$.
The left-right symmetry implies that, for the gauge couplings $g_{2L}$ and $g_{2R}$ of $SU(2)_L$ and $SU(2)_R$, $g_{2L} = g_{2R} = g_2$.

In the LRTH model proposed in \cite{lrth-review,0611015-su,lfv-lrth}, the global symmetry is $U(4) \times U(4)$
and the gauge subgroup is $SU(2)_L \times SU(2)_R \times U(1)_{B-L}$.
With the global symmetry $U(4) \times U(4)$ in the LRTH models,
the Higgs field and the twin Higgs in the fundamental representation of each $U(4)$
 can be written as $H$ = $( H_{L}, H_{R} )$ and $\hat{H}$ = $( \hat{H}_{L},\hat{H}_{R} )$, respectively.
After each Higgs develops a vacuum expectation value (VEV),
\begin{equation}
<H>^T = (0,0,0,f),\quad <\hat{H}>^T = (0,0,0,\hat{f}),
 \label{vevs}
\end{equation}
the global symmetry $U(4)\times U(4)$ breaks to $U(3)\times U(3)$, with the gauge group
$ SU(2)_L \times SU(2)_R \times U(1)_{B-L}$ down to the SM $U(1)_Y$.

After the Higgses obtain VEVs as shown in Eq. (\ref{vevs}), the breaking of the $SU(2)_R \times U(1)_{B-L}$
to $ U(1)_Y$ generates three massive gauge bosons, with masses proportional to $\sqrt{f^2 + \hat f^2}$.
The couplings of the these gauge bosons to the SM particles, so either precision measurements or direct searches greatly constrain their masses.
The masses of the these extra gauge bosons can be large enough to avoid the constraints from the electroweak precision
measurements by requiring $\hat f \gg f$. The problems of the large value of $\hat f $, however, can be eliminated
by imposing certain discrete symmetry which requires that the $\hat H$ is odd while all the other fields are even
so as to ensure the Higgs field $\hat H$ couples only to the gauge sector as described in ref.\cite{0611015-su}.

In such models, with the global symmetry breaking from ${\rm U}(4) \times{\rm U}(4)$ to ${\rm
U}(3)\times{\rm U}(3)$, and gauge symmetry from
${\rm SU}(2)_L\times {\rm SU}(2)_R \times {\rm U}(1)_{B-L}$ to
${\rm SU}(2)_L \times {\rm U}(1)_Y$ and  finally to ${\rm U}(1)_{EM}$, fourteen Goldstone bosons are generated, six of which are
eaten by the massive gauge bosons $Z_H$ and $W_H^{\pm}$ and the SM gauge bosons $Z^0$ and $W^{\pm}$, while the
rest of the Goldstone bosons contain the Higgses:
one neutral pseudoscalar, $\phi^0$, a pair of charged
scalars $\phi^\pm$, the SM physical Higgs $h$, and an ${\rm SU}(2)_L$ twin Higgs
doublet $\hat{h}=(\hat{h}_1^+, \hat{h}_2^0)$.

Since the effective Yukawa couplings suppressed by $f/\Lambda$,  with $\Lambda=4\pi \hat f$,
cannot account for the $\cal {O}(1)$ top Yukawa coupling.
To give the large top quark mass, vector-like quarks are introduced.
They also cancel the leading quadratic divergence of the SM gauge bosons and the top quark contributions
to the Higgs masses in the loop level, except for the new heavy gauge bosons, so the hierarchy problem settles down.
At the same time, the new particles such as the gauge bosons and the vector-like top singlet in the LRTH models
have rich phenomenology at the LHC\cite{lrth-pheno,0911.5567,lfv-lrth,0611015-su}.



Based on the Lagrangian given in Ref. \cite{0611015-su},
 we have, 
the couplings with fermions involved, which are concerned of our calculation in TABLE \ref{lrth-couplings},
\begin{table}[h]
\begin{tabular}{|c|c||c|c|}
\hline
particles & vertices & particles & vertices \\
\hline $W^+_{H\mu} \bar t b$ &  $ ~~e \gamma_{\mu} S_RP_R/( \sqrt{2} s_w) $ &
$W^{+\mu}_H \bar T b$ &    $ ~~e \gamma_{\mu} C_RP_L/( \sqrt{2} s_w)  $ \\
\hline $\Phi^+ \bar t b$ & $  -i (S_R m_b  P_L-y  S_L f P_R)/ f        $ &
$\Phi^+ \bar T b$ & $   ~~i (C_R m_b  P_L-y C_L f P_R)/f          $\\
\hline
\end{tabular}
\caption{The three-point couplings of  the charged gauge
boson-fermion-fermion and those of the scalar-fermion-fermion in the LRTH models.
The chirality projection operators are $P_{R, L}=(1\pm\gamma_5)/2$.
\label{lrth-couplings}  }
\end{table}

\hspace{-0.5cm}where the mixing angles are\cite{0611015-su}
\beq
S_L\sim sin\alpha_L \sim \frac{M} {m_T }sin x, ~~ S_R\sim sin\alpha_R \sim \frac{ M}{ m_T}(1 + sin2 x),~~ x = \frac{v}{\sqrt{2}f}.
\eeq
As for the parameter $M$ above, as we know, in the gauge invariant top Yukawa terms,
there is the mass mixing term $M \bar q_Lq_R$, allowed by gauge
invariance. $M\neq 0$ means there is mixing between the SM-like top quark and the heavy top quark.
The mixing parameter $M$ also be constrained by the $Z\to \bar b b$ branching ratio and oblique parameters
and it usually prefers to a small value\cite{0611015-su,0701071}.

Neutrino oscillations \cite{oscillneutrinos} imply that neutrinos are massive,
and the LRTH models try to explain the origin of the neutrino masses and mass hierarchy\cite{lfv-lrth}.
To provide lepton masses in the LRTH models, one can introduce three families doublets ${\rm SU}(2)_{L,R}$
which are charged under $SU(3)_c \times SU(2)_L \times SU(2)_R \times U(1)_{B-L}$ as
\begin{eqnarray}
    L_{L\alpha}=-i\left(\begin{array}{c}~\nu_{L\alpha}
\\l_{L\alpha}\end{array}\right) :({\bf 1},{\bf 2},{\bf 1},-1), \ \ \ \ \ &&
    ~~~L_{R\alpha}=\left(\begin{array}{c}~\nu_{R\alpha}
\\l_{R\alpha}\end{array}\right) :({\bf 1},{\bf 1},{\bf 2},-1),\nonumber
\end{eqnarray}
where the family index $\alpha$ runs from 1 to 3.

Leptons can acquire masses via non-renormalisable dimension 5 operators.
The charged leptons obtain their masses
via the following non-renormalisable dimension $5$ operators,
\begin{equation}
{y_l^{ij}\over \Lambda} (\bar{L}_{Li} H_L)(H_R^{\dagger}{L}_{Rj})+{y_{\nu}^{ij}\over \Lambda}
(\bar{L}_{L,i}\tau_2 H_L^*)(H_R^T\tau_2{L}_{Rj}) + {\rm {H.c.}} ,
\label{eq:Yukawalep}
\end{equation}
which will give rise to lepton Dirac mass terms $y_{\nu,l}^{ij} vf/\Lambda$, once $H_L$ and $H_R$ acquire VEVs.

The Majorana nature of the left- and right-handed neutrinos, however, makes one to induce
Majorana terms ( only the mass section) in dimension 5 operators,
\begin{equation}
{c_{L}\over \Lambda}\, \left( \overline{L}_{L\alpha} \tau_2 H_{L}^\dagger \right)^2+ {\rm H. c},
\qquad {c_{R} \over \Lambda}\, \left( \overline{L}_{R\alpha} \tau_2 H_{R}^\dagger \right)^2+ {\rm H. c. }~.
\label{MajoranaLR}\
\end{equation}
Once $H_L~(H_R)$ obtains a VEV, both neutrino chiralities obtain Majorana masses via
these operators, the smallness of the light neutrino masses,
however, can not be well explained.

However, if we assume that the twin Higgs $\hat{H}_R$ is forbidden to couple to the quarks to prevent the heavy top
quark from acquiring a large mass of order $y \hat f$, but it can couple to
the right-handed neutrinos, one may find that \cite{lfv-lrth}
\begin{equation}
{c_{\hat{H}} \over \Lambda}\, \left( \overline{L}_{R\alpha} \tau_2 \hat{H}_{R}^\dagger \right)^2+ {\rm H. c. }~,
\label{dimfiv}
\end{equation}
which will give a contribution to the Majorana mass of the heavy right-handed neutrino $\nu_R$,
in addition to those of  Eq.(\ref{MajoranaLR}).

So after the electroweak symmetry breaking, $H_R$ and $\hat{H}_R$ get VEVs, $f$ and $\hat{f}$ (Eq.(\ref{vevs})), respectively,
we can derive the following seesaw mass matrix for the LRTH model in the basis ($\nu_L$,$\nu_R$):
\begin{equation}
{\cal M} = \left(\begin{array}{cc}
c {v^ 2 \over 2\Lambda}  & y_{\nu} {v f \over \sqrt{2}\Lambda} \\
y_{\nu}^{T} {v f \over \sqrt{2}\Lambda} & c {f^ 2 \over \Lambda}+ c_{\hat{H}} {\hat{f}^ 2 \over \Lambda}
\end{array}\right)~.
\label{calM}
\end{equation}
In the one-generation case there is two massive states, a heavy ($\sim \nu_R$) and a light one.
For the case that $v <  f < \hat{f}$, the masses of the two eigenstates are about
$m_{\nu_{heavy}} \sim c_{\hat{H}} {\hat{f}^ 2 \over \Lambda}$ and $m_{\nu_{light}} = {c v^ 2 \over 2\Lambda} $ \cite{lfv-lrth}.

The Lagrangian in Eq.(\ref{eq:Yukawalep}), (\ref{MajoranaLR}), (\ref{dimfiv}) induces neutrino masses
and the mixings of different generation leptons, which may be a source of lepton flavour violating  \cite{lfv-lrth}.
In our case we will consider the contributions to the lepton flavour violating of the charged scalars, $\phi^\pm$
and the heavy gauge boson, $W_{H}$.
The relevant vertex interactions for these processes are explicated in the followings:
\begin{equation}
\phi^-\bar l \nu_{L,R}:\frac{i}{f}(m_{l_L,\nu_R} P_L-m_{\nu_{L},l_R}P_R) V_H\sim ic_H\frac{{\hat f}^2}{\Lambda f} V_H P_L,\label{hlv}
\end{equation}
\begin{equation}
W^-_{L,R}\bar l \nu_{L,R}:\frac{e}{\sqrt{2} s_w}\gamma_{\mu}   P_{L,R} V_H.
\label{wlv}
\end{equation}
where $V_H$ is the mixing matrix of the heavy neutrino and the leptons mediated by the charged scalars and
the heavy gauge bosons. The vertexes of $\phi^-\bar l \nu_{L,R}$ can also be expressed in the coupling constants. The
$\phi^-\bar l \nu_{R}$, for example, is also written as  $ic_H\frac{\hat f^2}{\Lambda f}P_L$
if we neglect the charged lepton masses and take $m_{\nu_h}=c_H\hat f^2/ \Lambda$.

\section{Muon $g-2$ anomaly and relevant constraints}\label{constraints}
\subsection{Numerical calculations}
In this paper, the light CP-even Higgs $h$ is taken as the SM-like Higgs, $m_h=$ 125 GeV.
Since the muon $g-2$ anomaly favors a light charged pseudoscalar with a large coupling to lepton and a heavy right-handed neutrinos,
we scan over $m_\phi$ and $m_{\nu_R}$ in the following ranges\cite{lfv-lrth,0611015-su}:
\beq
100 ~{\rm GeV} <m_{\phi^\pm}< 1000~ {\rm GeV},~~5000~GeV<m_{\nu_R}<50000~GeV.
\eeq

In the following calculation, the following constraints are considered:
\begin{itemize}
\item[(1)] From theoretical constraints and precision electroweak data,
The theoretical constraints such as those from the unitarity and coupling-constant perturbativity, and the constraints from
the oblique parameters $S$, $T$, $U$ will be considered\cite{1809.05857}.

\item[(2)] From the lepton number violating signals of the top partners\cite{0911.5567,lfv-lrth}, we can see that
right-handed neutrinos prefer to have a very large mass and the charged scalars are heavy.
we can also have the constrains $m_{v_R}> m_{T}$ and  $m_{T}> m_{W_H}$\cite{0911.5567}.

\item[(3)] The constraints from the signal data of the 125 GeV Higgs will be important, since the couplings of the 125 GeV Higgs with the SM particles in LRTH model can deviate from the SM ones and the SM-like decay modes may be modified severely.
Moreover, when $m_{\phi^0}$ is smaller than $m_h/2=62.5$ GeV,
the decay $h\to \phi^0\phi^0$ is kinematically allowed, and the experimental data of the 125 GeV Higgs will constrain it.
We will perform $\chi^2_h$ calculation for the signal strengths
of the 125 GeV Higgs, which will be discussed detailedly in Sec. IV.

\item[(4)] $f$ and $M$ parameter: 
  The indirect constraints on $f$ come from the $Z$-pole precision measurements, the low energy
neutral current process and the high energy precision measurements off the Z-pole: all these data
prefer the parameter f to be larger than 500-600 GeV \cite{0611015-su}. On
the other hand, it cannot be too large since the fine tuning is more severe for large $f$.

\hspace{0.5cm} In the LRTH, furthermore, the mass of the top partner $T$ is determined by the given values of $f$ and $M$.
Currently, the masses of the new heavy particles, such as the $T$ have been constrained by the LHC experiments,
as described in Refs. \cite{LHC-expr-1,LHC-expr-2}.
In other words, the LHC data also imply some indirect constraints on the allowed ranges of both the
parameters f and M through their correlations with $m_T$, as discussed in Ref. \cite{f-mt}. For
example, the top partner $T$ with mass below 656 GeV are excluded at $95\%$ confidence level according
to the ATLAS data \cite{ATLAS-mT} if one takes the assumption of a branching ratio $BR(T \to W^+b) = 1$.

By taking the above constraints from the electroweak precision measurements and the LHC data into account,
we here assume that the values of the parameter $f$ and $M$ are in the ranges of
\beq
 500GeV \leq f \leq 1500GeV, \quad 0 \leq M \leq 500GeV,
\eeq
in our numerical evaluations.

\end{itemize}

\subsection{ Muon $g-2$ in the LRTH models }

\begin{figure}[tb]
  \epsfig{file=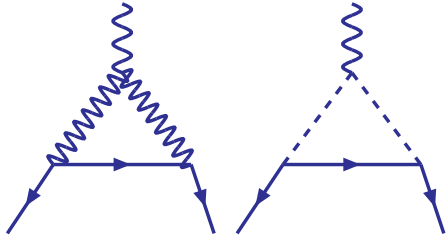,height=3.3cm}
\vspace{-0.5cm} \caption{ The one-loop contributions to $a_\mu$ in the LRTH models.}
\label{fig1}
\end{figure}

 In the LRTH, the muon $g-2$ contributions are obtained via the
one-loop diagrams induced by the Higgs bosons and also from the
two-loop Barr-Zee diagrams mediated by $\phi^0$, $h$ and $\phi^\pm$.

the one-loop contributions are give in the following, and the corresponding
figures are given in Fig.\ref{fig1}.
We can write down them one by one\cite{1loop-deltaamu}:
 \beq
    \Delta a_\mu^{\mbox{$\scriptscriptstyle{\rm LRTH}$}}({\rm 1loop})_W =
    \frac{e^2}{2 s_W^2} \frac{ m_{\mu}^2}{8 \pi^2 } \int_0^1 dx \frac{-6x^3-2x^2}{m_{W_H}^2 x+ m_{v_R}^2(1-x)},
\label{1loopW}
\eeq
 \beq
    \Delta a_\mu^{\mbox{$\scriptscriptstyle{\rm LRTH}$}}({\rm 1loop})_H =
    \frac{m_{\nu_R}^2}{f^2} \frac{ m_{\mu}^2}{8 \pi^2 } \int_0^1 dx \frac{2(x^3-x^2)}{m_H^2 x+ m_{v_R}^2(1-x)},
\label{1loopH}
\eeq

Before we immerse into the two-loop contribution, we discuss the coupling between the boson and the scalars,
and we find that they all vanish, if we parameterize the scalars in the Goldstone
bosons fields as\cite{0611015-su},
\begin{eqnarray}
\label{unitary}
    \begin{array}{ll}
  N~ \to  \frac{\sqrt{2}\hat{f}}{F(\cos x+2\frac{\sin
x}{x})}\phi^0, &
~\hat{N}~ \to  -\frac{\sqrt{2}f\cos x}{3F}\phi^0, \\
  h_1\to  0, &
~h_2 \to
\frac{v+h}{\sqrt{2}}-i\frac{x\hat{f}}{\sqrt{2}F(\cos x+2\frac{\sin
x}{x})}\phi^0, \\
  C~ \to  -\frac{x\hat{f}}{F\sin x} \phi^+, &
~\hat{C}~ \to  \frac{{f\cos x}}{F} \phi^+. \\
\end{array}
\end{eqnarray}
where the $N, ~ \hat{N},~ h_1,~h_2,C,~\hat{C}$ are in the Goldstone
bosons fields,
\begin{equation}
\label{GBrep}
    H= i\frac{\sin\sqrt{\chi}}{\sqrt{\chi}}e^{i\frac{N}{2f}}\left(%
\begin{array}{c}
  h_1 \\ h_2\\
  C\\
  N-if\sqrt{\chi}\cot\sqrt{\chi}\\
\end{array}%
\right), \ \ \hat{H} =
i\frac{\sin\sqrt{\hat{\chi}}}{\sqrt{\hat{\chi}}}
e^{i\frac{\hat{N}}{2\hat{f}}}\left(%
\begin{array}{c}
  \hat{h}_1 \\ \hat{h}_2\\
  \hat{C}\\
  \hat{N}-i\hat{f}\sqrt{\hat{\chi}}\cot\sqrt{\hat{\chi}}\\
\end{array}%
\right).
\end{equation}

By this parameterization, the requirement of vanishing gauge-Higgs
mixing terms can be satisfied, i.e, in this redefinition of the
Higgs fields, the couplings $WZ\phi^+$, $W\gamma\phi^+$, $WW\phi^0$,
$WZ_H\phi^+$, $W\gamma_H\phi^+$, $W\phi^0\phi^+$, and $Wh\phi^+$ are
zero, which has been
verified and is quite different with those in other models such as the littlest Higgs models\cite{ppwpi-1302.1840}.

\begin{figure}[tb]
  \epsfig{file=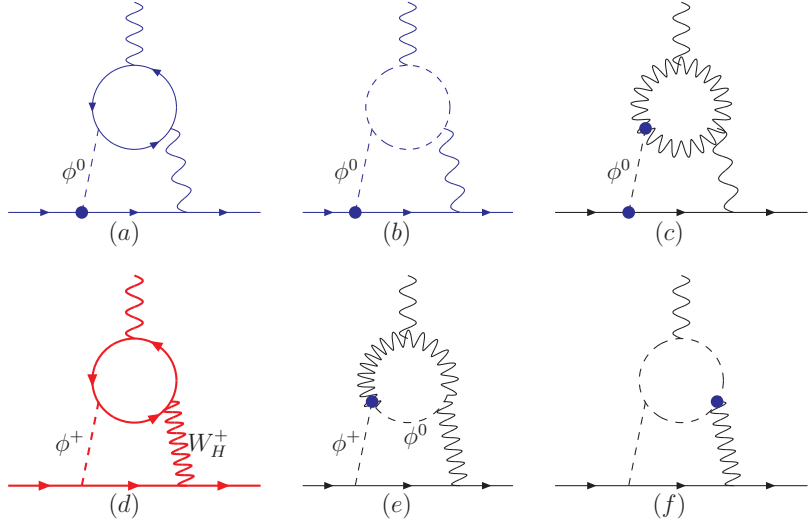,height=7.3cm}
\vspace{-0.5cm} \caption{ The potential two-loop contributions to $a_\mu$ the LRTH models.}
\label{fig2}
\end{figure}

Since the coupling between the boson and the scalars $W\gamma\phi^+$, $Wh\phi^+$ and $W\phi^0\phi^+$ have been
vanished, so the Barr-Zee 2-loop diagrams (e) (f) (c) in Fig. (\ref{fig2})  disappear.
Barr-Zee 2-loop diagrams (a) (b) in Fig. (\ref{fig2}) may not be negligible even though
the vertexes such as $\phi^0 \bar\mu \mu $ is very small, which is
proportional to the muon mass, much smaller than the masses the top and heavy top in our special case.
%
So there are (a) (b) (d) left contributing to $a_\mu$.

We can write down the Barr-Zee two-loop contributions of the diagram (a) (b) (d) respectively\cite{g-2-cal,1502.04199,1507.07567},
\begin{align}
	\Delta a_\mu^{(a)} =  -\frac{ 4m^2_\mu}{e}\frac{-e^3}{128 \pi^4 } \sum_{f_j=t,T}  \frac{N_f^c Q^2_f}{m_\mu}  \sum_{i=h,\phi^0}  \Gamma_{\ell_f\ell_i}^i \Gamma_{f_jf_j}^i  \frac{m^{f_j}}{m_i^2} g_i^{(a)}(r_{f_j}^i) ,
\end{align}
where $N_f^c$ and $Q_f$ are the number of colours and charge of fermion $f$, respectively,
$\Gamma_{f_j f_i}^i$s are the couplings of the scalars to the fermions,  and $r_f^i \equiv m_f^2/m_\mu^2$.
The loop function is given by
\begin{align}
	g_i^{(a)}(r) = \int_0^1 \dd x\, \frac{N_i(x)}{x(1-x) - r} \ln\left( \frac{x(1-x)}{r} \right) ,
\end{align}
where
\begin{align}
	N_h(x) = 2x(1-x) - 1 \,, \quad N_{\phi^0}(x) = -1 \,.
\end{align}

\begin{align}
\begin{split}
	\Delta a_\mu^{(b)} &=-\frac{ 4m^2_\mu}{e} \frac{ e^3}{128\sqrt2 \pi^4} \frac{v}{m_\mu} \quad\sum_{i=h,\phi^0} \frac{\Gamma_{\ell_f\ell_i}^i}{m_i^2} \zeta^i \lambda_{H^+H^-H_i^0} g_i^{(b)}\left( \frac{m_{H^+}^2}{m_i^2} \right)  ,
	\end{split}
\end{align}
where $\zeta^h = -\zeta^H = -\zeta^A = 1$ and the loop function is
\begin{align}
	g_{h,H,A}^{(b)}(r) = \int_0^1 \dd x\, \frac{x(1-x)}{x(1-x) - r} \ln\left( \frac{r}{x(1-x)} \right) .
\end{align}

\bea
   \Delta a_\mu^{(d)}({ttb+bbt})&=&
   -\frac{ 4m^2_\mu}{e} \frac{-e^3S_RV_H}{1024\pi^4 sin^2\theta_w} \frac{N_t^c V_{tb}^*}{m_\phi^2-m_{W_H}^2} \\ && \nn
 \int_0^1 dx[Q_tx+Q_b(1-x)]
    \left[ G\left( \frac{m_t^2}{m_{H^+}^2}, \frac{m_b^2}{m_{H^+}^2} \right) - G\left( \frac{m_t^2}{m_W^2}, \frac{m_b^2}{m_W^2} \right) \right]  \\ && \nn
      \times \left[ \left( {\Gamma_{tb}^{\phi^+,L}}^* {\Gamma_{\nu_f\mu}^{\phi^+}} \right) \frac{m_b}{m_{\mu}} x(1-x) - \left( {\Gamma_{tb}^{\phi^+,R}}^* {\Gamma_{\nu_f\mu}^{\phi^+}} \right) \frac{m_t}{m_{\mu}} x(1+x) \right] \\
       \Delta a_\mu^{(d)}({TTb+bbT})&=&
   -\frac{ 4m^2_\mu}{e} \frac{-e^3C_RV_H}{1024\pi^4 sin^2\theta_w} \frac{1}{m_\phi^2-m_{W_H}^2} \\ && \nn
 \int_0^1 dx[Q_Tx+Q_b(1-x)]
    \left[ G\left( \frac{m_T^2}{m_{H^+}^2}, \frac{m_b^2}{m_{H^+}^2} \right) - G\left( \frac{m_T^2}{m_W^2}, \frac{m_b^2}{m_W^2} \right) \right]  \\ && \nn
      \times \left[ \left( {\Gamma_{Tb}^{\phi^+,R}}^* {\Gamma_{\nu_f\mu}^{\phi^+}} \right) \frac{m_b}{m_{\mu}} x(1-x) - \left( {\Gamma_{Tb}^{\phi^+,L}}^* {\Gamma_{\nu_f\mu}^{\phi^+}} \right) \frac{m_T}{m_{\mu}} x(1+x) \right]
\label{barr-zee}
\eea
where the loop function is defined as,
\begin{align}
	G(r^a, r^b) = \frac{\ln\left( \frac{r^a x + r^b(1-x)}{x(1-x)} \right)}{x(1-x) - r^a x - r^b(1-x)} \,,
\end{align}
and $\Gamma_{tb}^{\phi^+,R}$ and $\Gamma_{tb}^{\phi^+,L} $ are the right-handed and left-handed couplings of the vertex $\phi^+\bar t b$,
which are given in Table \ref{lrth-couplings}. From Table \ref{lrth-couplings}, we also see that the top vector-like partner $T$ enter into the triangle loop just as the top quark, and the contribution to $a_\mu$ is
 \beq
 \Delta a_\mu^{TTb}({\rm 2loop-BZ})= \Delta a_\mu^{ttb}({\rm 2loop-BZ})(m_t \to m_T,~N_t^c \to N_T^c)
 \eeq
where for the vector-like fermion, $N_T^c=1$.

By the way, we should note that in the Barr-Zee 2-loop diagrams there are no two scalars or two $W^\pm$ charged bosons connect to the
triangle loop simultaneously, which is induced by the helicity constraints since between the two charged particles, the fermion is the bottom
qurak, which mass is much smaller than that of the top quark, and the slash momentum terms must vanish undergoing a single $\gamma$ matrix.
Of course, the discussion here is very crudely, and explicit and detailed discussion can be found in Ref. \cite{1502.04199}.

As the enhancement factor $m^2_f/m^2_\mu$ could easily
overcome the loop suppression phase space factor $\alpha/\pi$, the two-loop
contributions can be larger than one-loop ones. In the LRTH, since
the CP-odd Higgs coupling to the lepton is proportional to
$m_{v_R}$, the LRTH can sizably enhance the muon g-2 for a light
CP-odd scalar and a large right-handed neutrino mass $m_{v_R}$.

\section{Global fit of the 125 GeV Higgs}


The 125 GeV Higgs signal data include a large number of observales and we will perform a global fit to the 125 GeV Higgs signal data.
For the given neutral SM-like scalar-field $h$ and its couplings, the $\chi^2_h$ function can be defined as
\begin{align}
\chi^2_h\; =\; \sum_k\; \frac{\left(\mu_k  - \hat{\mu}_k\right)^2}{\sigma_k^2}\, ,
\end{align}
where $k$ runs over the different production(decay) channels considered, and $\hat{\mu}_k$ and $\sigma_k$ denote the measured Higgs signal strengths
 and their one-sigma errors, respectively.
$\mu_k$ is the corresponding theoretical predictions for the LRTH parameters, as given later in Eqs.~(\ref{eq:ratios}) and (\ref{ratios2}).


\subsection{Relevant Lagrangian and the Couplings}
After diagonaling, the Yukawa lagrangians can be written as,
\beq\label{yukawa}
 \mathcal L_Y  =   - \sum_{f=d,l} y_f h\bar{f} P_R f  - \sum_{f=t,T} y_f h\bar{u} P_L u
\eeq

From Eq.(\ref{yukawa}) and the couplings in Ref. \cite{0611015-su}, we can get the interactions between the Higgs boson
and the pairs of $\bar b b,~ \bar l l,~t\bar{t},~T\overline{T},~VV(V=W,W_H),\phi^+\phi^-$:
\bea
y_b=-\frac{m_{b}}{v}C_{L}C_{R}=-\frac{m_{b}}{v}\rho_b,~ y_l =-\frac{m_{l}}{v}C_{L}C_{R}=-\frac{m_{l}}{v}\rho_l,&&\\
~y_{t}=-\frac{m_{t}}{v}C_{L}C_{R} =-\frac{m_{t}}{v}\rho_t,~y_{T}=y(S_{R}S_{L}-C_{L}C_{R}x)/\sqrt{2},&& \\
h~W^+_{\mu}~ W^-_{\nu} : ~~e m_{W}/s_W= \rho_W  m_W^2/v,~~ \rho_W=ev/(m_w s_W)     && \\
h~W_{H\mu}^+~ W_{H\nu}^-: -e^2f~ x~ g_{\mu\nu}/(\sqrt{2} s_w^2)  =y_{W_H} g_{\mu\nu},   &&\\
h \phi^+\phi^-:y_\phi, ~y_\phi= -x\frac{2m_h^2-2m_\phi^2}{3\sqrt{2}f}. &&
\eea
$\rho_t=\rho_b=\rho_\tau=C_{L}C_{R},~\rho_W=\frac{e*v}{m_WS_W} $ are the ratios of $hff,~hWW$ vertexes in LRTH and the standard models.
\subsection{Higgs Signal Strengths}

The so-called signal strengths, which are employed in the experimental data on Higgs searches,
measuring the observable cross sections in ratio to the corresponding SM predictions.
At the LHC, the SM-like Higgs particle is generated by the following
relevant production mechanisms: gluon fusion ($ g g \rightarrow H$),
vector boson fusion ($ q q^{\prime} \rightarrow q q^{\prime} VV\rightarrow q q^{\prime} H$),
associated production with a vector boson ($q \bar q^{\prime} \rightarrow W H/Z H $),
and the associated production with a $t \bar t$ pair  ($q \bar q/ gg \rightarrow t \bar t  H$).
The Higgs decay channels are $\gamma \gamma$, $Z Z^{(*)}$, $W W^{(*)}$, $b \bar b$ and $\tau^+ \tau^-$.

In order to fit the experimental measurements, we can write down the following ratios:
\begin{align}
\mu_{gg\gamma\gamma} &\equiv
\frac{\sigma(pp\to h)\, \text{Br} (h\to \gamma\gamma)}{\sigma(pp\to H)_{\mathrm{SM}}\, \text{Br} (H\to \gamma\gamma)_{\mathrm{SM}}}\, ,
\qquad &
\mu_{ t\bar th\gamma\gamma} &\equiv
\frac{\sigma(pp\to t\bar th)\, \text{Br} (h\to \gamma\gamma)}{\sigma(pp\to t\bar tH)_{\mathrm{SM}}\, \text{Br} (H\to \gamma\gamma)_{\mathrm{SM}}}\, ,
\notag \\[7pt]
\mu_{ggVV} &\equiv
\frac{\sigma(pp\to h)\, \text{Br} (h\to VV)}{\sigma(pp\to H)_{\mathrm{SM}}\, \text{Br} (H\to VV)_{\mathrm{SM}}}\, ,
&
\mu_{t\bar t hVV} &\equiv
\frac{\sigma(pp\to t\bar th)\, \text{Br}(h\to VV)}{\sigma(pp\to t\bar tH)_{\mathrm{SM}}\, \text{Br}(H\to VV)_{\mathrm{SM}}}\, ,
\notag \\[7pt]
\mu_{ggff} &\equiv
\frac{\sigma(pp\to h)\, \text{Br} (h\to ff)}{\sigma(pp\to H)_{\mathrm{SM}}\, \text{Br} (H\to ff)_{\mathrm{SM}}}\, ,
&
\mu_{t\bar t hff} &\equiv
\frac{\sigma(pp\to t\bar th)\, \text{Br}(h\to ff)}{\sigma(pp\to t\bar tH)_{\mathrm{SM}}\, \text{Br}(H\to ff)_{\mathrm{SM}}}\, ,
\notag \\[7pt]
\mu_{Vh\gamma\gamma} &\equiv
\frac{\sigma(pp\to Vh)\,\text{Br}(h\to \gamma\gamma)}{\sigma(pp\to VH)_{\mathrm{SM}}\,\text{Br}(H\to \gamma\gamma)_{\mathrm{SM}}}\, ,
\qquad &
\mu_{ VBF\gamma\gamma} &\equiv
\frac{\sigma(pp\to qqh)\, \text{Br} (h\to \gamma\gamma)}{\sigma(pp\to q qH)_{\mathrm{SM}}\, \text{Br} (H\to \gamma\gamma)_{\mathrm{SM}}}\, ,
\notag \\[7pt]
\mu_{VhVV} &\equiv
\frac{\sigma(pp\to Vh)\, \text{Br} (h\to VV)}{\sigma(pp\to H)_{\mathrm{SM}}\, \text{Br} (H\to VV)_{\mathrm{SM}}}\, ,
&
\mu_{VBFVV} &\equiv
\frac{\sigma(pp\to qqh)\, \text{Br}(h\to VV)}{\sigma(pp\to qqH)_{\mathrm{SM}}\, \text{Br}(H\to VV)_{\mathrm{SM}}}\, ,
\notag \\[7pt]
\mu_{Vhff} &\equiv
\frac{\sigma(pp\to Vh)\, \text{Br} (h\to ff)}{\sigma(pp\to H)_{\mathrm{SM}}\, \text{Br} (H\to ff)_{\mathrm{SM}}}\, ,
&
\mu_{VBFff} &\equiv
\frac{\sigma(pp\to qqh)\, \text{Br}(h\to ff)}{\sigma(pp\to qqH)_{\mathrm{SM}}\, \text{Br}(H\to ff)_{\mathrm{SM}}}\, ,
\label{eq:ratios}
\end{align}
where $V={W, \!\ Z}$.

The ratio of the branching fraction will be expressed as:
\beq \label{Brratios}
\frac{\text{Br}(h \to X )}{\text{Br}(H\to X)_{\mathrm{SM}}}\; =\;
\dfrac{1}{\rho(h)}\;  \frac{\Gamma(h \to X )}{\Gamma(H\to X)_{\mathrm{SM}}}
\;\, ,
\eeq
where $\rho(h)$ is the total decay width of the scalar $h$ in units of the SM Higgs width,
\bea \\
 \rho(h) &=& \frac{ \Gamma (h)}{\Gamma_{\mathrm{SM}} (H) }\\
         &=& \frac{\Gamma^{BSM}(h)+\Gamma(h\to \varphi^0\varphi^0) }{ \Gamma_{\mathrm{SM}} (H) }\\
         &=& \frac{\Gamma^{BSM}(h) }{ \Gamma_{\mathrm{SM}} (H) } + \frac{\Gamma(h\to \varphi^0\varphi^0) }{ \Gamma_{\mathrm{SM}} (H)},
\eea
where the existence of the $h\to \varphi^0\varphi^0$ means in the LRTH models when $\varphi^0$ mass is less than $m_h/2$,
 the channel $h\to \varphi^0\varphi^0 $ will be open, and the
total width of $h$ should changed into $\Gamma^{LRTH}(h)+\Gamma(h\to \varphi^0\varphi^0)$, where $\Gamma^{LRTH}(h)$ is
corresponding to SM channels. $\Gamma(h\to \varphi^0\varphi^0)$ can be written as
\beq
\Gamma(h\to \varphi^0\varphi^0)=\frac{g^2_{h\varphi^0\varphi^0}}{8\pi m_h} \sqrt{1-\frac{4 m^2_\varphi}{m_h^2}}
\eeq
where $g_{h\varphi^0\varphi^0}=\frac{vm_h^2}{54 f^2}[11+15(1-\frac{2m_\varphi^2}{m_h^2})]$\cite{1312.4004}.

Particularizing to the LRTH and assuming only one dominant production channel in each case, we have:
\begin{align}  \nm
\mu_{gg\gamma\gamma}= C_{gg}C_{\gamma\gamma} \rho(h)^{-1},~~\mu_{ggVV}= C_{gg}\rho_W^2\rho(h)^{-1},~~ \mu_{ggff}= C_{gg}\rho_f^2 \rho(h)^{-1},&& \\ \nm
\mu_{t\bar t h\gamma\gamma}= \rho_t^2 C_{\gamma\gamma} \rho(h)^{-1},~~\mu_{t\bar t hVV}= \rho_t^2\rho_W^2\rho(h)^{-1}, ~~\mu_{t\bar t hff}= \rho_t^2\rho_f \rho(h)^{-1},&& \\  \nm
\mu_{VBF\gamma\gamma}= \rho_W^2 C_{\gamma\gamma} \rho(h)^{-1},~~\mu_{VBFVV}= \rho_W^2\rho_W^2\rho(h)^{-1},~~ \mu_{VBFff}= \rho_W^2\rho_f^2 \rho(h)^{-1},&& \\
\mu_{Vh\gamma\gamma}= \rho_W^2 C_{\gamma\gamma} \rho(h)^{-1},~~\mu_{VhVV}= \rho_W^2\rho_W^2\rho(h)^{-1}, ~~\mu_{Vhff}= \rho_W^2\rho_f \rho(h)^{-1}
\label{ratios2}
\end{align}
Note that $ \rho_W= \rho_Z$. 


The one-loop functions are given by
\beq  \label{gluonfusion}
C_{gg} = \frac{\sigma(gg\to h)}{\sigma(gg\to h)_{\mathrm{SM}}} =
\frac{ | \sum_{q=t,T} y_q \mathcal{F}(x_q) |^2 }{ |\sum_{q=t} y_t \mathcal{F}(x_q) |^2}
\eeq
where $y_t=y_t v/\sqrt{2}$, and
\begin{align}\label{gammascaling}
C_{\gamma\gamma} \; &=\; \frac{\Gamma(h\to\gamma\gamma)}{\Gamma(h\to\gamma\gamma)_{\mathrm{SM}}}
\\[5pt] & =\;
\frac{\Big|\sum_f y_f N_C^f  Q_f^2   \mathcal{F}(x_f) + \mathcal{F}_1(x_W)y_W + \mathcal{F}_1(x_{W_H})y_{W_H}+ \mathcal{F}_0(x_{\phi^\pm}) y_\phi \Big|^2 }
{\Big|\sum_f    N_C^f  Q_f^2  \mathcal{F}(x_f) + \mathcal{G}(x_W)  \Big|^2} \qquad
\notag\end{align}
with $N_C^f$ and $Q_f$ the number of colours and the electric charge of the fermion $f$, and $x_f = 4 m_f^2/M_{h}^2$,
 $x_W = 4 M_W^2/M_{h}^2$ and $x_{\phi^\pm} = 4 M_{\phi^\pm}^2/M_{h}^2$.
Note that the ratios \eqn{eq:ratios} are defined for $M_{h} = M_{h_{\mathrm{SM}}}$.
%
The functions $\mathcal{F}(x_f)$ and $\mathcal{F}_1(x_W)$ contain the contributions of the triangular 1-loop from fermions and $W^\pm$ bosons.
The masses of the first two fermion generations will be neglected.
Since $\mathcal{F}(x_f)$ vanishes for massless fermions, we only need to consider the top 
and the vector-like top contributions, which correspond large Yukawa couplings.

The explicit expressions of the different loop functions can be given as:
\begin{align}  \label{functions}
\mathcal{F}(x)\; &=\; \frac{x}{2} [4+(x-1)f(x)]\, , \qquad &
\mathcal{F}_1(x)\; &=\; -2 -3x + \Big(\frac{3}{2}x - \frac{3}{4}x^2\Big)f(x)\, ,
\notag \\
\mathcal{F}_0(x)\; &=\; -x - \frac{x^2}{4}f(x)\, ,  &
 \mathcal{K}(x) \; &= \;  - \frac{x}{2} f(x)   \,,
\end{align}
with
\begin{equation}
f(x)\; =\; \begin{cases} -4\arcsin^2(1/\sqrt{x})\, , \quad & x\geqslant1 \\[3pt] \Big[\ln\Big( \frac{1+\sqrt{1-x}}{1-\sqrt{1-x}}\Big)- i\pi \Big]^2\, , & x<1 \end{cases} \, .
\end{equation}

\section{Results and discussions}
\begin{figure}[tb]
\epsfig{file=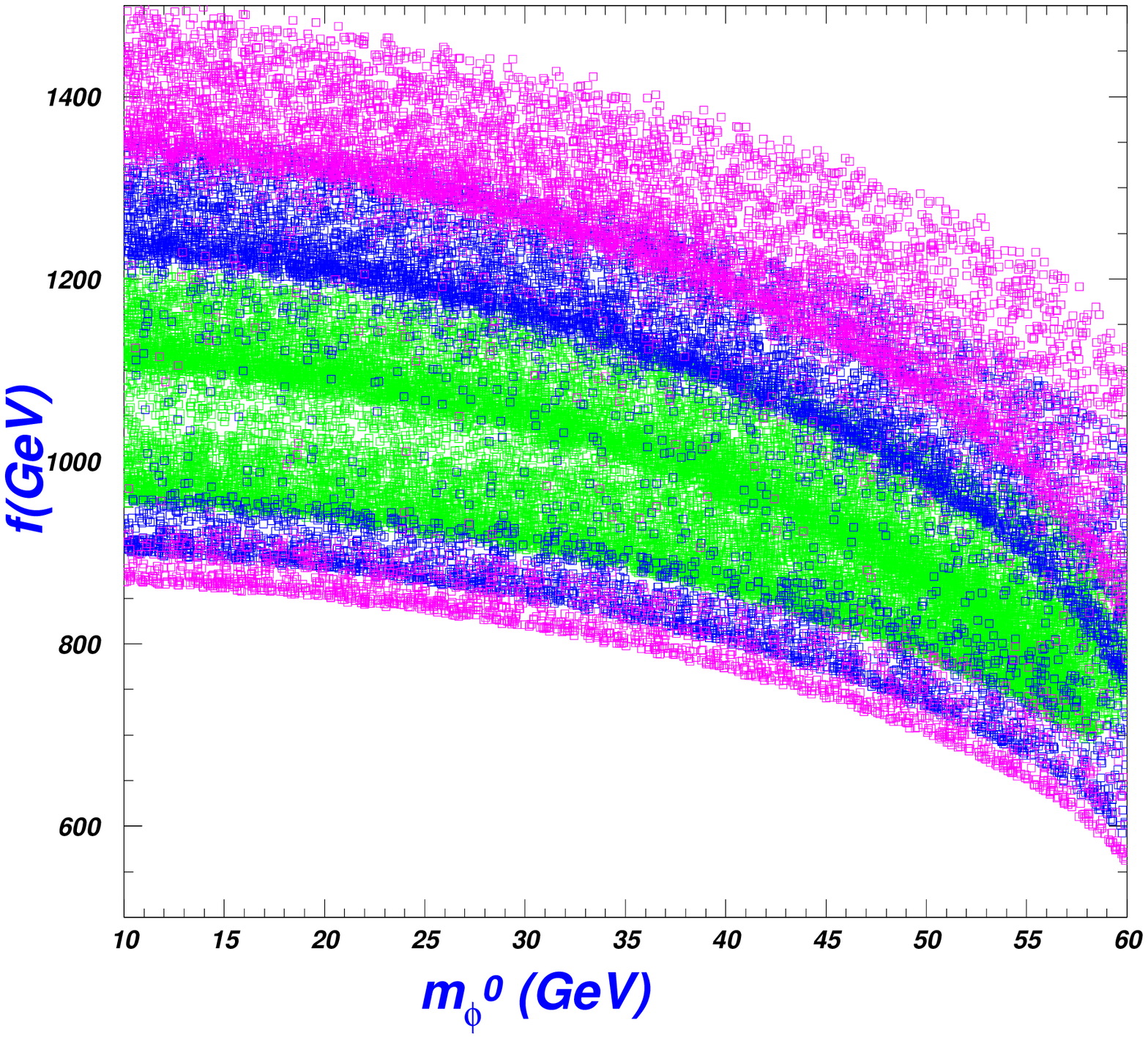,height=5.7cm}
\epsfig{file=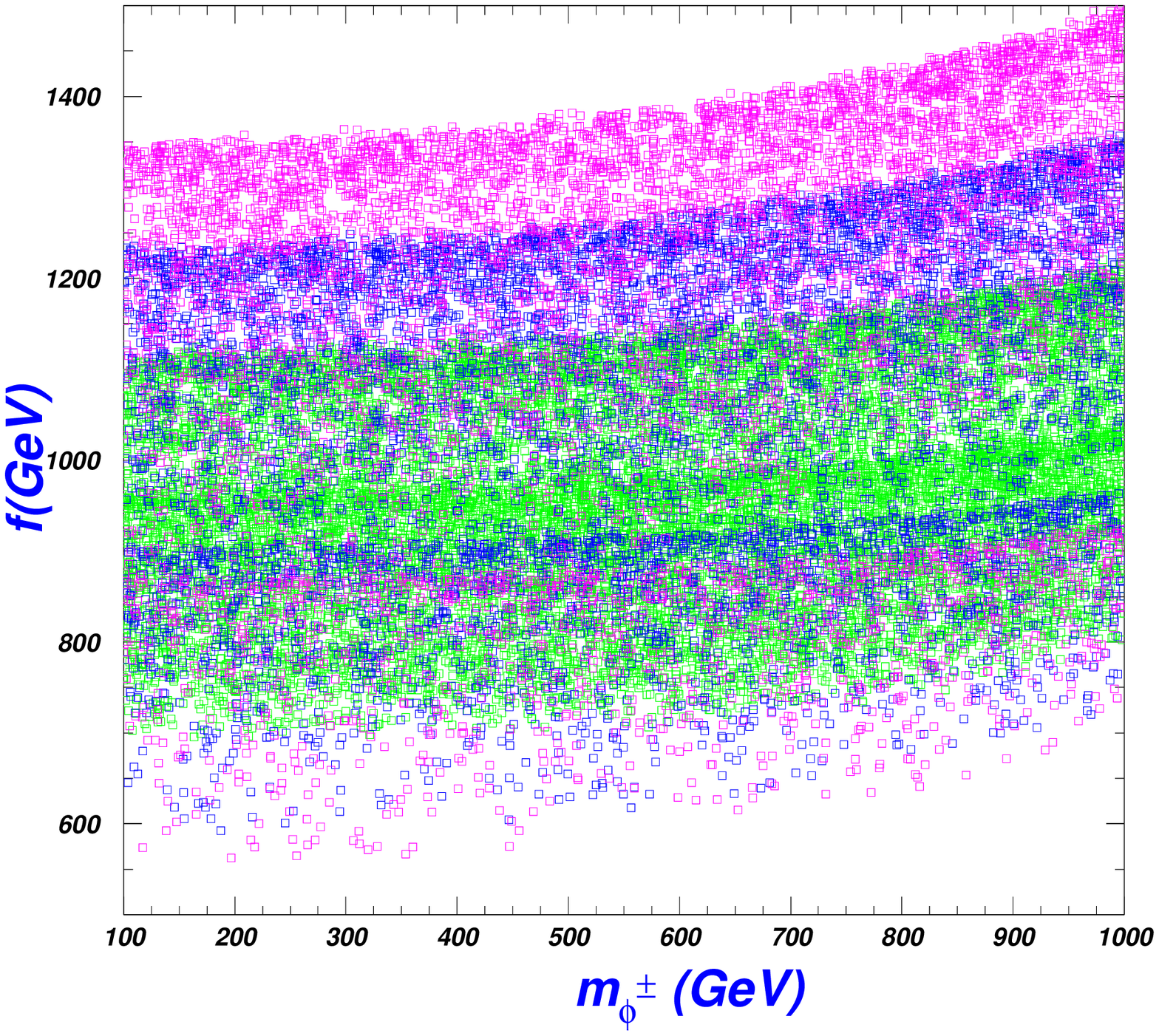,height=5.7cm}
\epsfig{file=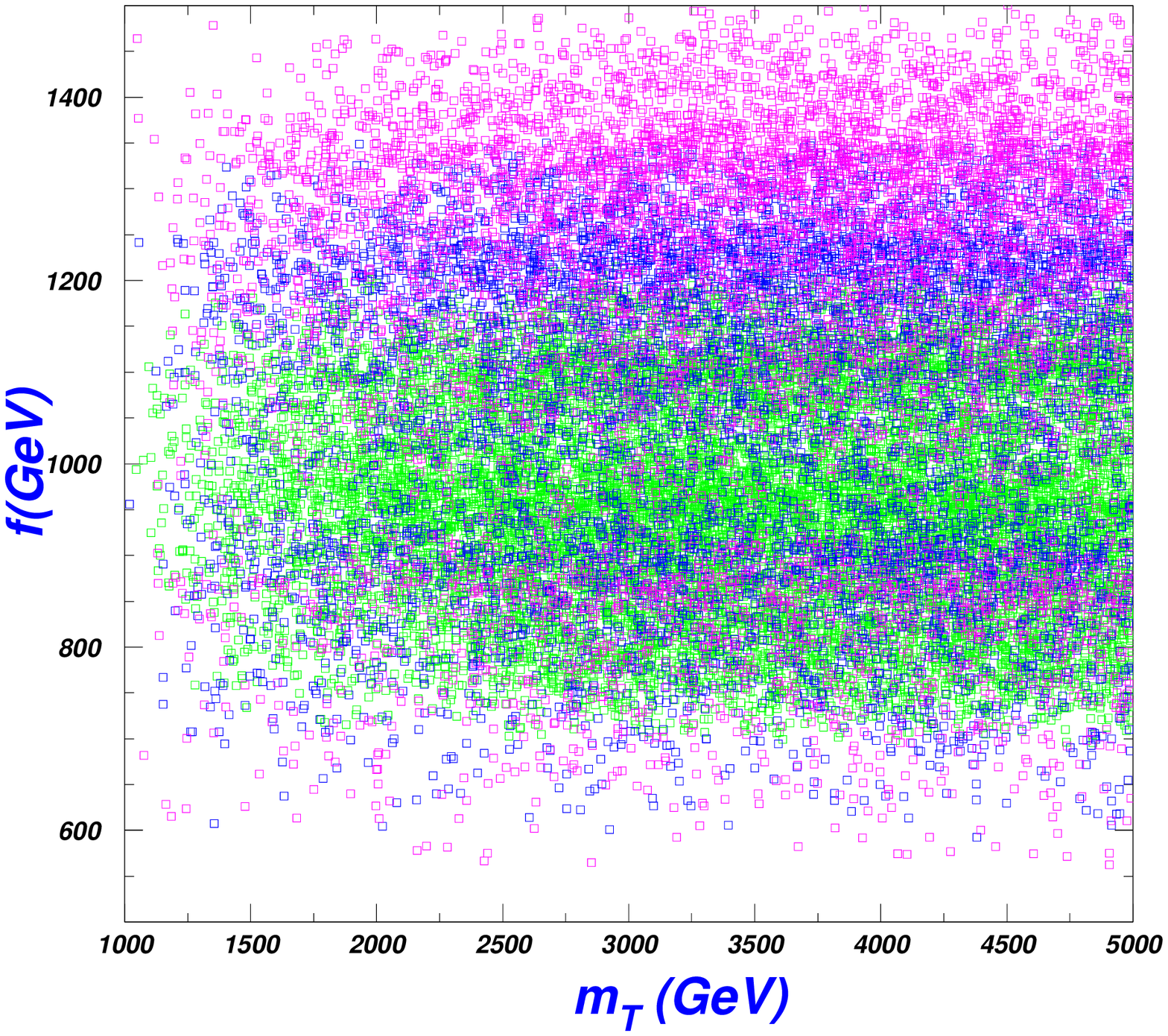,height=5.7cm}
\epsfig{file=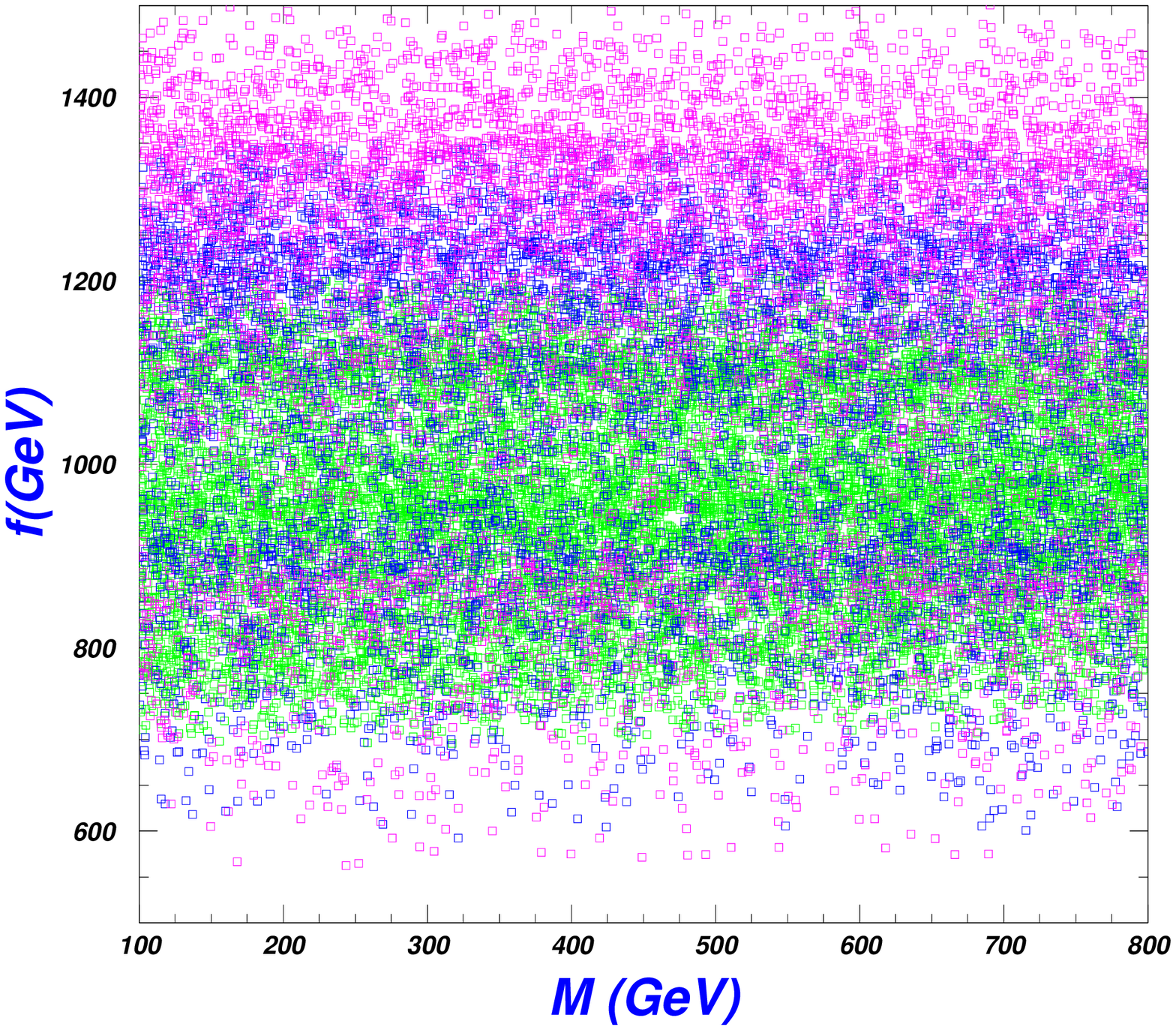,height=5.7cm}
\vspace{-0.5cm} \caption{The surviving samples
within 1$\sigma$, 2$\sigma$, and 3$\sigma$
ranges of $\chi^2_h$ on the planes
of $f$ versus $m_{\phi^0}$, $m_{\phi^\pm}$,~$m_T$, and $M$. The green, blue
and the pink points are respectively within the $1\sigma$,$2\sigma$, and $3\sigma$ regions of $\chi^2_h$.
}
\label{chi2}
\end{figure}

\begin{figure}[tb]
\epsfig{file=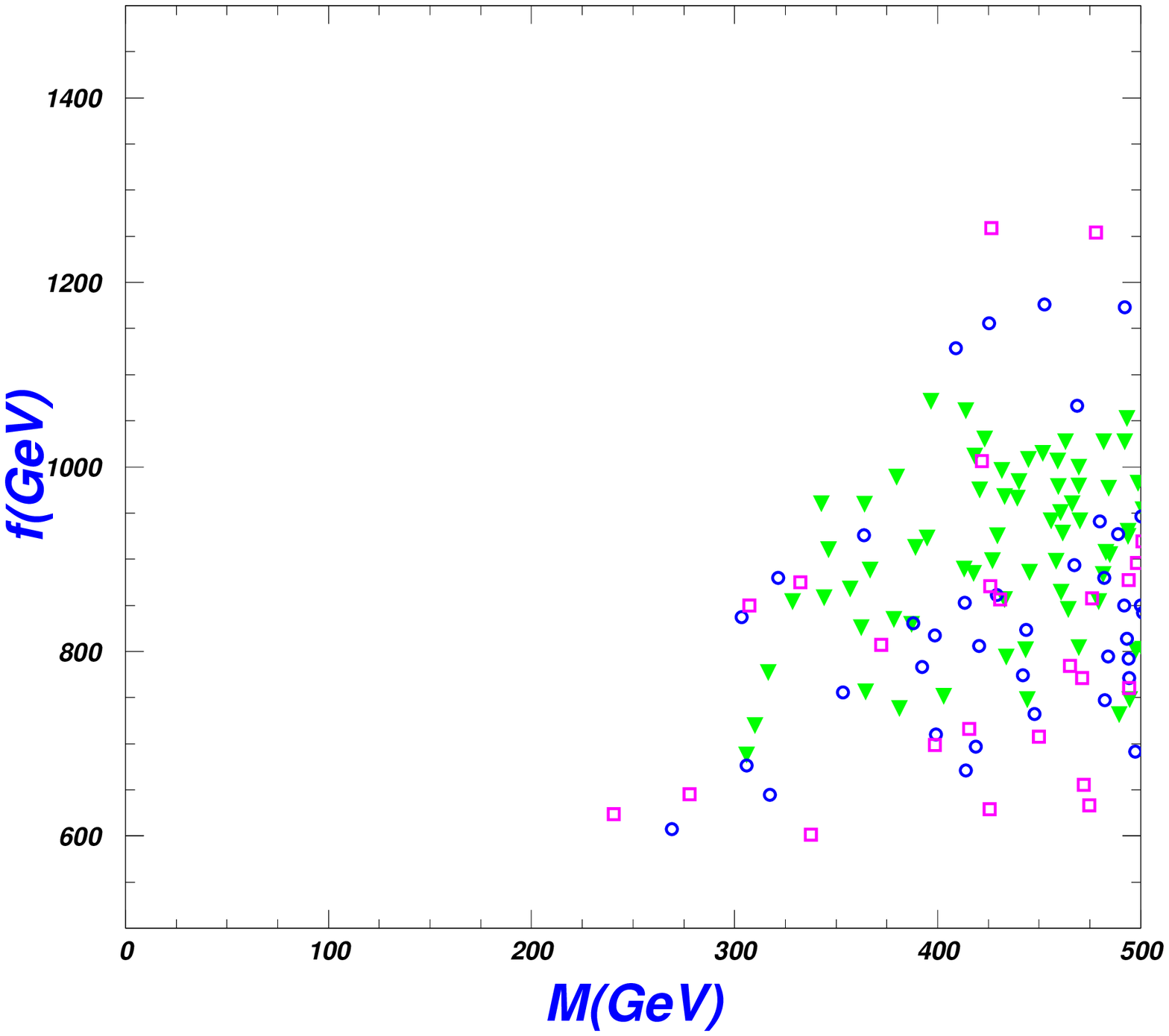,height=5.7cm}
\epsfig{file=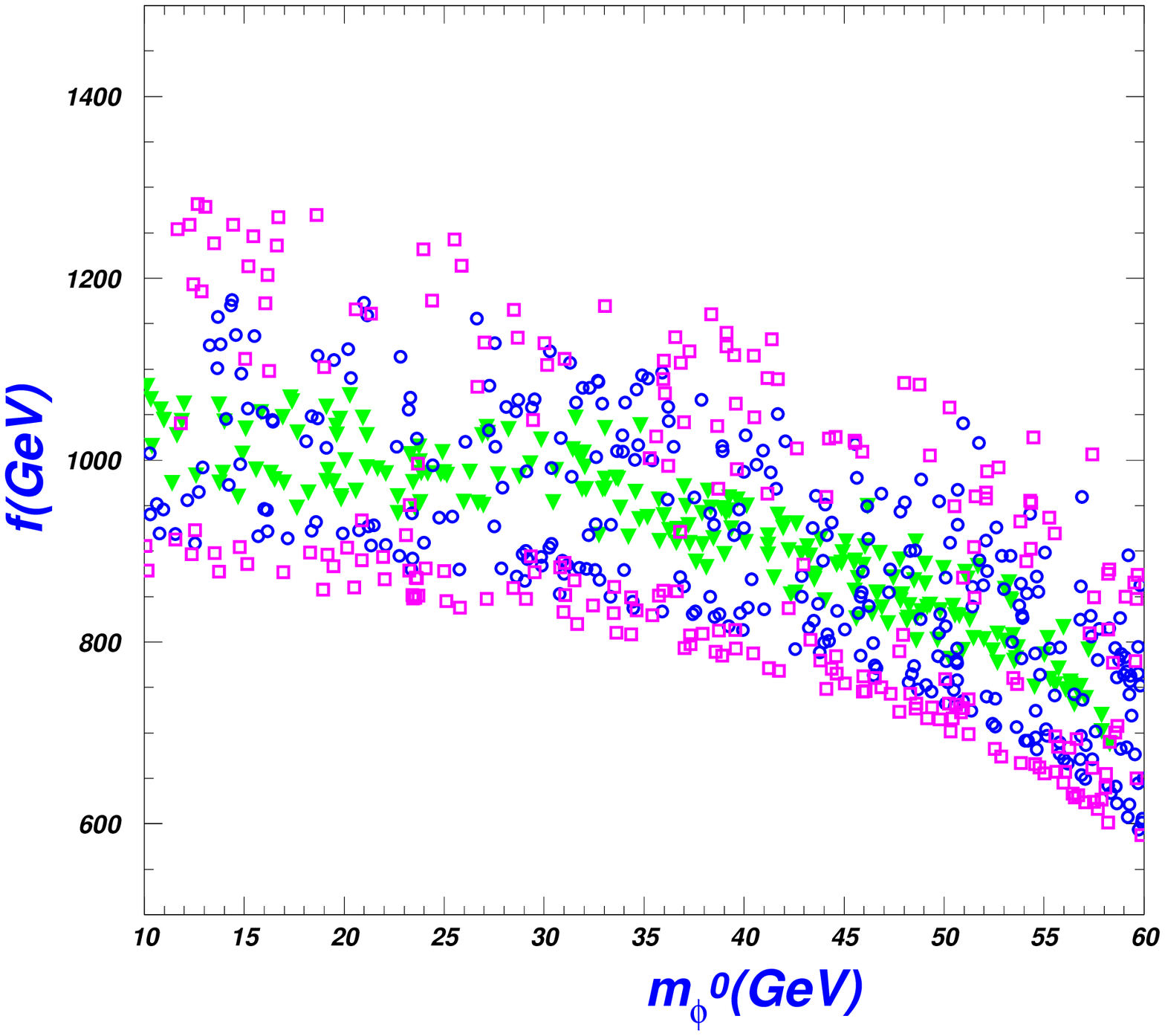,height=5.7cm}
\epsfig{file=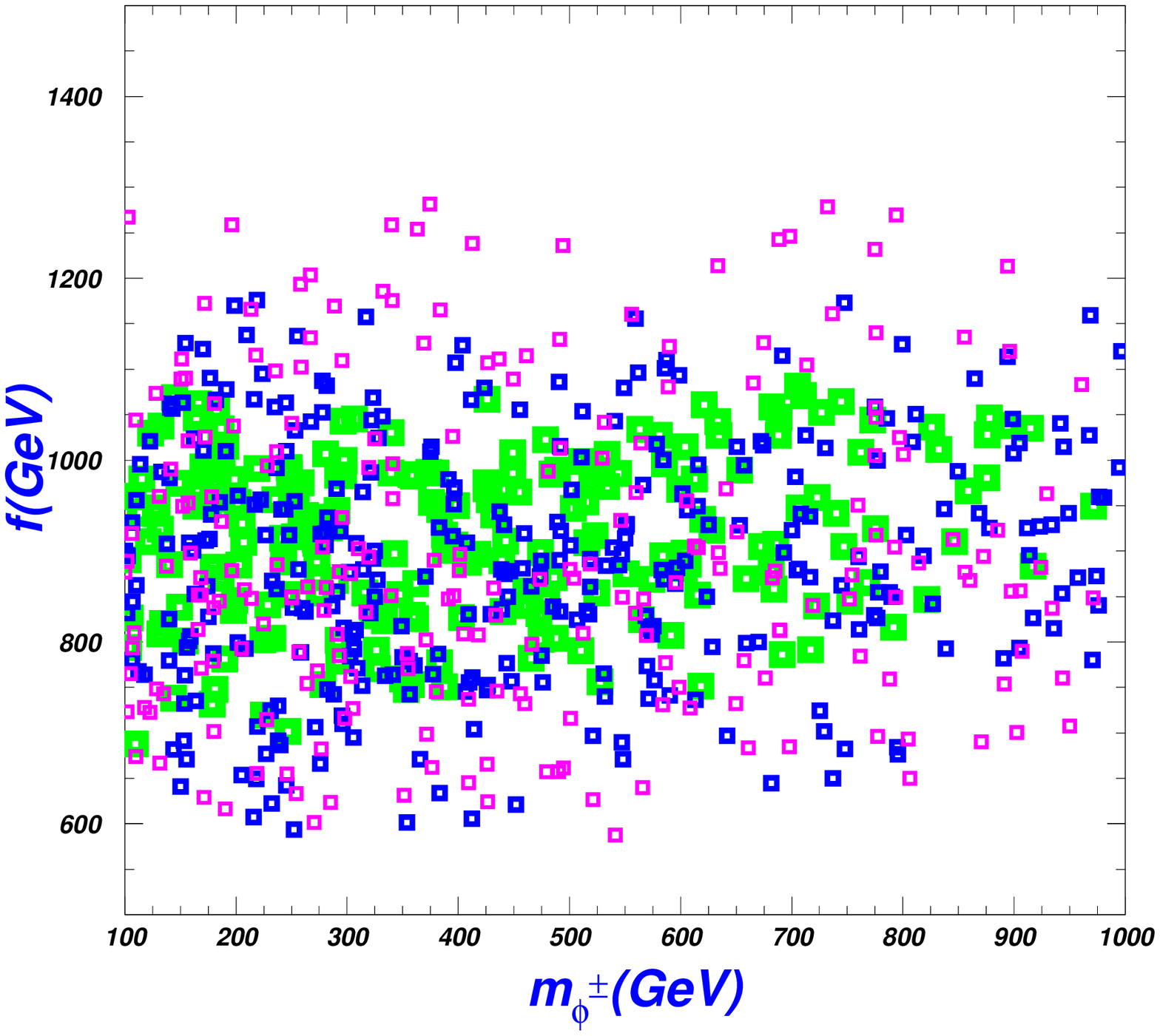,height=5.7cm}
\epsfig{file=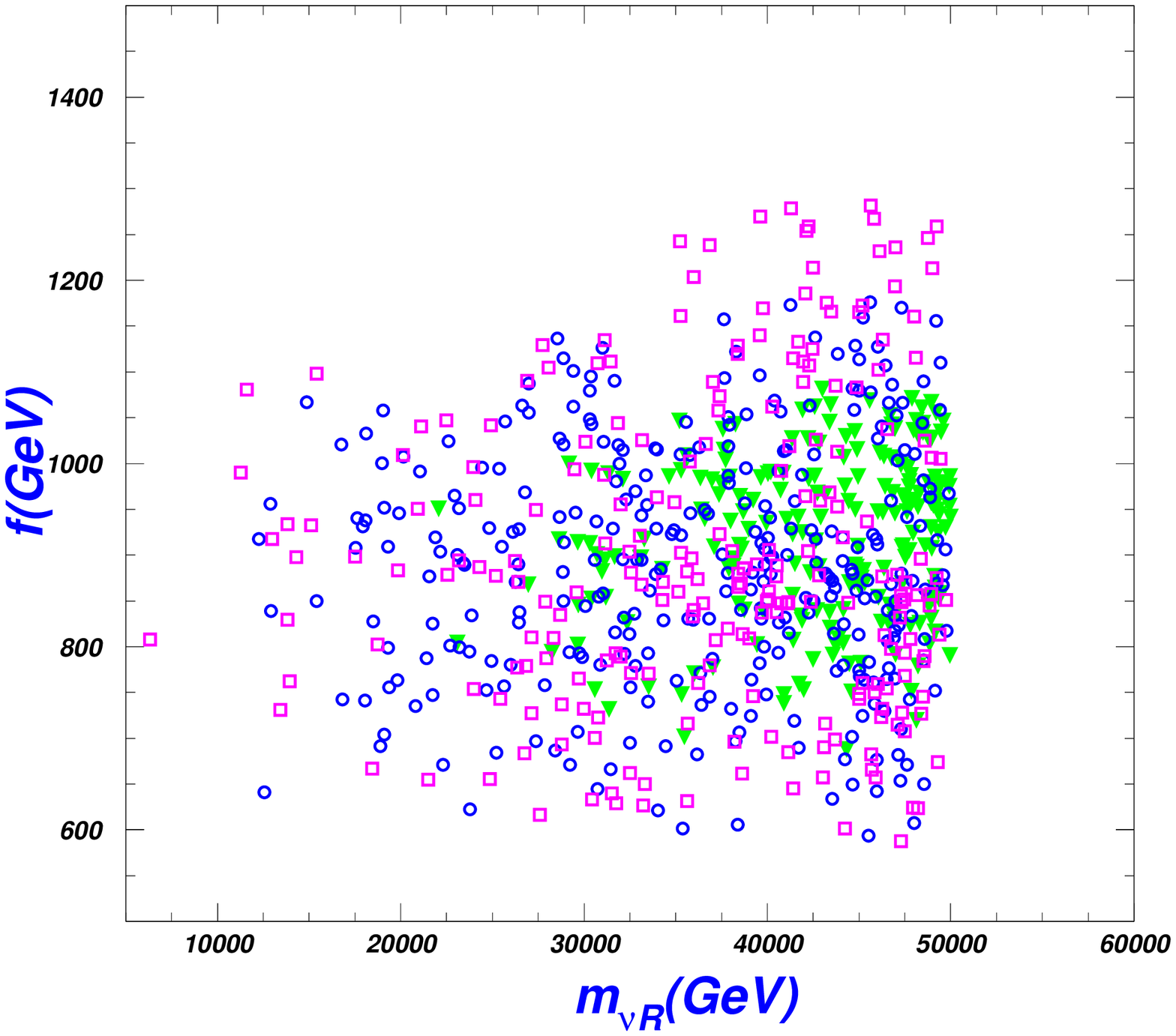,height=5.7cm}
\vspace{-0.25cm} \caption{
The samples satisfying the constraints of Higgs global fit
within 1$\sigma$, 2$\sigma$, and 3$\sigma$ ranges, on the planes
of $f$ versus $M$, $m_{\phi^0}$, $m_{\phi^\pm}$, and $m_{\nu_R}$,
with the constraints of the $a_{\mu}$ from the experiemnts.
 All the samples are allowed by the constraints of muon $g-2$.
 The green, blue
and the pink points are respectively within the $1\sigma$,$2\sigma$, and $3\sigma$ regions of $\chi^2_h$.
}
\label{mug2}
\end{figure}
In Fig. \ref{chi2}, we project the surviving samples
within 1$\sigma$, 2$\sigma$, and 3$\sigma$
ranges of $\chi^2_h$ on the planes
of $f$ versus $m_{\phi^0}$, $m_{\phi^\pm}$,~$m_T$, and $M$,
the exclusion limits from searches for Higgs at LEP,
the signal data of the 125 GeV Higgs, and the flavor changing constraints of $\mu \to e\gamma$\cite{lfv-lrth}.
Fig. \ref{chi2} shows that the of $\chi^2_h$ value favors a little large $f$.
We can see from the upper panel that if the value of $f$ is small, the value of $\chi^2_h$
prefer to have a large $m_{\phi^0}$ and a small $m_{\phi^\pm}$.
From the lower-left panel of Fig. \ref{chi2} that the value of $\chi^2_h$ is favored to
a large top partner mass $m_T$.

In Fig. \ref{mug2}, we project the surviving samples within 1$\sigma$, 2$\sigma$, and 3$\sigma$
on the planes of $f$ versus $m_{\phi^0}$, $m_{\phi^\pm}$,~$m_T$, and $M$
 after imposing the constraints from the muon g-2 anomaly,
the lepton flavor changing decay.
The lower-left panel shows that the surviving data preferring to a large mixing parameter $M$,
about $200-500$ GeV.

Fig. \ref{mug2} shows that with the limits from muon $g-2$, the Higgs global fit and the
lepton decay $\mu\to e\gamma$ being satisfied,
the muon $g-2$ anomaly can be explained in the regions of 200 GeV $\leq M\leq $ 500 GeV,
700 GeV $\leq f\leq $ 1100 GeV,
10 GeV $\leq m_{\phi^0}\leq $ 60 GeV,  100 GeV $\leq m_{\phi^\pm}\leq $ 900 GeV, and $m_{\nu_R}\geq$ 15 TeV.
Fig. \ref{mug2} shows that
in the range of 10 GeV $\leq m_{\phi^0}\leq $ 60 GeV and a light $f$ constrained by the decay $\mu\to e\gamma$,
the muon $g-2$ anomaly can be explained for a large enough $m_{\nu_R}$, which constraint severely the models which introduce
extra right-handed neutrinos to give the natural light neutrino masses.
Since the contributions of $m_{\nu_R}$ to the muon $g-2$ anomaly have destructive
interference with the prediction, there may not exist many samples, and the model survives in narrow space,
as shown in the lower-right panel of Fig. \ref{mug2}.


\section{Conclusion}
The muon $g-2$ anomaly can be explained in the LRTH model.
After imposing various relevant theoretical and experimental constraints,
we performed a scan over the parameter space of this model to identify the ranges in
favor of the muon $g-2$ explanation, and the Higgs direct search limits from LHC constraint strongly.
We find that the muon g-2 anomaly can be accommodated in the region of 300 GeV $\leq M\leq $ 500 GeV,
700 GeV $\leq f\leq $ 1100 GeV,
10 GeV $\leq m_{\phi^0}\leq $ 60 GeV,  100 GeV $\leq m_{\phi^\pm}\leq $ 900 GeV,
and $m_{\nu_R}\geq$ 15 TeV, after imposing the joint constraints from
the theory, the precision electroweak data, the 125 GeV Higgs signal data, and the leptonic decay.

\section*{Acknowledgment}
The author would appreciate the helpful discussions with Fei Wang, Wen-Yu Wang and Lei Wang.
This work was supported by the National Natural Science Foundation
of China under grant 11675147, 11605110 and
by the Academic Improvement Project of Zhengzhou University. 




\end{document}